\documentclass[bibnotes,a4paper,twocolumn,prl]{revtex4}%
\usepackage{amsfonts}
\usepackage{amsmath}
\usepackage{amssymb}
\usepackage{graphicx}%
\graphicspath{{Z:/SUPRA/fluctuations_fflo/article_PRB/}}
\ifx\pdfoutput\relax\let\pdfoutput=\undefined\fi
\newcount\msipdfoutput
\ifx\pdfoutput\undefined\else
\ifcase\pdfoutput\else
\ifx\paperwidth\undefined\else
\ifdim\paperheight=0pt\relax\else\pdfpageheight\paperheight\fi
\ifdim\paperwidth=0pt\relax\else\pdfpagewidth\paperwidth\fi
\fi\fi\fi
\begin{document}
\title{Magnetic moment manipulation by a Josephson current}
\author{F. K\textsc{onschelle}}
\email{f.konschelle@cpmoh.u-bordeaux1.fr}
\author{A. B\textsc{uzdin}}
\altaffiliation{Also at \emph{Institut Universitaire de France}.\ }

\affiliation{Condensed Matter Theory Group, CPMOH, Universit\'{e} de Bordeaux and CNRS.
F-33405 Talence, France}
\date{\today, Document published in Phys. Rev. Lett. \textbf{102}, 017001 (2009).}
\startpage{1}

\begin{abstract}
We consider a Josephson junction where the weak-link is formed by a
non-centrosymmetric ferromagnet. In such a junction, the superconducting
current acts as a direct driving force on the magnetic moment. We show that
the a.c. Josephson effect generates a magnetic precession providing then a
feedback to the current. Magnetic dynamics result in several anomalies of
current-phase relations (second harmonic, dissipative current) which are
strongly enhanced near the ferromagnetic resonance frequency.

\end{abstract}
\maketitle

\bigskip

Many interesting phenomena have been observed recently in the field of
spintronics: the spin-dependent electric current and inversely the
current-dependent magnetization orientation (see for example
\cite{zutic_fabian_dassarma_RMP.76.323.2004,hauptmann_paaske_lindelof.2008}).
Moreover, it is well known that spin-orbit interaction may be of primary
importance for spintronic, namely for systems using a two-dimensional electron
gas \cite{b.winkler.2003}. In the superconductor/ferromagnet/superconductor
(S/F/S) Josephson junctions, the spin-orbit interaction in a ferromagnet
without inversion symmetry provides a mechanism for a direct (linear) coupling
between the magnetic moment and the superconducting current
\cite{buzdin:107005.2008}. Similar anomalous properties have been predicted
for Josephson junctions with spin-polarized quantum point contact in a two
dimensional electron gas \cite{reynoso_etal:107001.2008}. S/F/S junctions are
known to reveal a transition to $\pi$-phase, where the superconducting phase
difference $\varphi$ in the ground state is equal to $\pi$ \cite{buzdin(2005)}%
. However, the current-phase relation (CPR) in such a $\pi$-junction has a
usual sinusoidal form $I=I_{c}\sin\varphi$, where the critical current $I_{c}$
depends in a damped oscillatory manner on the modulus of the ferromagnet
exchange field. In a non-centrosymmetric ferromagnetic junction, called
hereafter $\varphi_{0}$-junction, the time reversal symmetry is broken and the
CPR becomes $I=I_{c}\sin\left(  \varphi-\varphi_{0}\right)  $, where the phase
shift $\varphi_{0}$ is proportional to the magnetic moment perpendicular to
the gradient of the asymmetric spin-orbit potential \cite{buzdin:107005.2008}.
Therefore, manipulation of the internal magnetic moment can be achieved via
the superconducting phase difference (\emph{i.e. }by Josephson current).

In the present work we study theoretically the spin dynamics associated with
such $\varphi_{0}$-junctions. Though there is a lot of experimental progress
in studying the static properties of S/F/S junctions, little is known about
the spin-dynamics in S/F systems. Note here the pioneering work
\cite{bell_Milikisyants_Huber_Aarts:047002.2008} where a sharpening of the
ferromagnetic resonance was observed below the superconducting transition in
Nb/Ni$_{80}$Fe$_{20}$ system. Theoretically, the single spin dynamics
interplay with a Josephson effect has been studied in
\cite{zhu_balatsky.2003,bulaevskii_hruska_shnirman_etal:177001.2004,zhu_nussinov_shnirman_balatsky.2004,nussinov_shnirman_arovas_balatsky_zhu:214520.2005}%
. More recently, the dynamically induced triplet proximity effect in S/F/S
junctions was studied in
\cite{takahashi_hikino_mori:057003.2007,houzet:057009.2008}, while the
junctions with composite regions (including several F regions with different
magnetization) were discussed in
\cite{Waintal_Brouwer.65.054407.2002,braude:207001.2008}. Here we consider a
simple S/F/S $\varphi_{0}$-junction in a low frequency regime $\hslash
\omega_{J}\ll T_{c}$ ($\omega_{J}=2eV/\hslash$ being the Josephson angular
frequency \cite{b.likharev.1986}), which allows us to use the quasi-static
approach to treat the superconducting subsystem in contrast with the case
analyzed in \cite{takahashi_hikino_mori:057003.2007,houzet:057009.2008}. We
demonstrate that a d.c. superconducting current may produce a strong
orientation effect on the F layer magnetic moment. More interestingly, the
a.c. Josephson effect, \emph{i.e.} applying a d.c. voltage $V$ to the
$\varphi_{0}$-junction, would produce current oscillations and consequently
magnetic precession. This precession may be monitored by the appearance of
higher harmonics in CPR as well as a d.c. component of the current. In
particular regimes, a total reversal of the magnetization could be
observed.\ In the case of strong coupling between magnetic and superconducting
subsystems, complicated non-linear dynamic regimes emerge.\ 

\begin{figure}[ptb]
\begin{center}
\includegraphics[width=3.4in,angle=0]{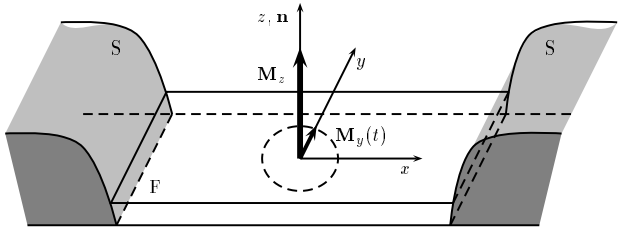}
\end{center}
\caption{Geometry of the considered $\varphi_{0}$-junction. The ferromagnetic
easy-axis is directed along the $z$-axis, which is also the direction
$\mathbf{n}$ of the gradient of the spin-orbit potential. The magnetization
component $\mathbf{M}_{y}$ is coupled with Josephson current through the phase
shift term $\varphi_{0}\propto\mathbf{n.}\left(  \mathbf{M\wedge\nabla}%
\Psi\right)  $, where $\Psi$ is the superconducting order parameter
($\mathbf{\nabla}\Psi$ is along the $x$-axis in the system considered here).}%
\label{FIG_schema}%
\end{figure}

To demonstrate the unusual properties of the $\varphi_{0}$-junction, we
consider the case of an easy-axis magnetic anisotropy of the F material (see
Fig.\ref{FIG_schema}). Both the easy axis and gradient of the asymmetric
spin-orbit potential $\mathbf{n}$ are along the $z$-axis. Note that suitable
candidates for the F interlayer may be MnSi or FeGe.\ In these systems, the
lack of inversion center comes from the crystalline structure, but the origin
of broken-inversion symmetry may also be extrinsic, like in a situation near
the surface of a thin F film. In the following, we completely disregard the
magnetic induction. Indeed the magnetic induction in the $\left(  xy\right)  $
plane is negligible for the thin F layer considered in this paper, whereas the
demagnetization factor cancels the internal induction along the $z$-axis
$\left(  N=1\right)  $. The coupling between F and S subsystems due to the
orbital effect has been studied in \cite{hikino_mori.2008} and it appears to
be very weak, and quadratic over magnetic moment $\mathbf{M}$ for the case
when the flux of $\mathbf{M}$ through the F layer is small in comparison with
flux quantum $\Phi_{0}=h/2e$.

The superconducting part of the energy of a $\varphi_{0}$-junction is
\begin{equation}
E_{s}\left(  \varphi,\varphi_{0}\right)  =E_{J}\left[  1-\cos\left(
\varphi-\varphi_{0}\right)  \right]  ,
\end{equation}
where $E_{J}=\Phi_{0}I_{c}/2\pi$ is the Josephson energy, $I_{c}$ is the
critical current and $\varphi_{0}$ is proportional to the $M_{y}$ component of
the magnetic moment (see Fig.\ref{FIG_schema}). Therefore, when the magnetic
moment is oriented along the $z$-axis, we have the usual Josephson junction
with $\varphi_{0}=0$. Assuming the ballistic limit we may estimate the
characteristic Josephson energy as \cite{buzdin(2005)} $\Phi_{0}I_{c}/S\sim
T_{c}k_{F}^{2}\sin\ell/\ell$ with $\ell=4hL/\hslash v_{F}$, where $S,L$ and
$h$ are the section, the length and the exchange field of the F layer,
respectively. The phase shift is \
\begin{equation}
\varphi_{0}=\ell\frac{v_{\text{so}}}{v_{F}}\frac{M_{y}}{M_{0}}%
\end{equation}
where the parameter $v_{\text{so}}/v_{F}$ characterizes the relative strength
of the spin-orbit interaction \cite{buzdin:107005.2008}. Further on we assume
that $v_{\text{so}}/v_{F}\sim0.1$. If the temperature is well below the Curie
temperature, $M_{0}=\left\Vert \mathbf{M}\right\Vert $ can be considered as a
constant equal to the saturation magnetization of the F layer. The magnetic
energy contribution is reduced to the anisotropy energy
\begin{equation}
E_{M}=-\frac{K\mathcal{V}}{2}\left(  \frac{M_{z}}{M_{0}}\right)  ^{2},
\end{equation}
where $K$ is an anisotropy constant and $\mathcal{V}$ is the volume of the F
layer.

Naturally, we may expect that the most interesting situation corresponds to
the case when the magnetic anisotropy energy does not exceed too much the
Josephson energy. From the measurements
\cite{rusanov_hesselberth_aarts_buzdin:057002.2004} on permalloy with very
weak anisotropy, we may estimate $K\sim4.10^{-5}%
\operatorname{K}%
.%
\operatorname{\text{\AA}}%
^{-3}$. On the other hand, typical value of $L$ in S/F/S junction is $L\sim10%
\operatorname{nm}%
$ and $\sin\ell/\ell\sim1$.\ Then, the ratio of the Josephson over magnetic
energy would be $E_{J}/E_{M}\sim100$ for $T_{c}\sim10%
\operatorname{K}%
$. Naturally, in the more realistic case of stronger anisotropy this ratio
would be smaller but it is plausible to expect a great variety of regimes from
$E_{J}/E_{M}\ll1$ to $E_{J}/E_{M}\gg1$. \ \ 

Let us now \ consider the case when a constant current $I<I_{c}$ is applied to
the $\varphi_{0}$-junction. The total energy is (see, \emph{e.g.
}\cite{b.likharev.1986}):
\begin{equation}
E_{\text{tot}}=-\frac{\Phi_{0}}{2\pi}\varphi I+E_{s}\left(  \varphi
,\varphi_{0}\right)  +E_{M}\left(  \varphi_{0}\right)  , \label{EQ_energy}%
\end{equation}
and both the superconducting phase shift difference $\varphi$ and the rotation
of the magnetic moment $M_{y}=M_{0}\sin\theta$ (where $\theta$ is the angle
between the $z$-axis and the direction of $\mathbf{M}$) are determined from
the energy minimum conditions $\partial_{\varphi}E_{\text{tot}}=\partial
_{\varphi_{0}}E_{\text{tot}}=0$. It results in
\begin{equation}
\sin\theta=\frac{I}{I_{c}}\Gamma\text{ \ with }\Gamma=\frac{E_{J}%
}{K\mathcal{V}}\ell\frac{v_{\text{so}}}{v_{F}}, \label{EQ_phi_0_current_bias}%
\end{equation}
which signifies that a superconducting current provokes the rotation of the
magnetic moment $M_{y}$ in the $\left(  yz\right)  $ plane. Therefore, for
small values of the rotation, $\theta\left(  I\right)  $ dependence is linear.
In principle, the parameter $\Gamma$ can be larger than one. In that case,
when the condition $I/I_{c}\geq1/\Gamma$ is fulfilled, the magnetic moment
will be oriented along the $y$-axis. Therefore, applying a d.c.
superconducting current switches the direction of the magnetization, whereas
applying an a.c. current on a $\varphi_{0}$-junction could generate the
precession of the magnetic moment.

We briefly comment on the situation when the direction of the gradient of the
spin-orbit potential is perpendicular (along $y$) to the easy axis $z$. To
consider this case we simply need to take $\varphi_{0}=\ell\left(
v_{\text{so}}/v_{F}\right)  \cos\theta$. The total energy $(\ref{EQ_energy})$
has two minima $\theta=\left(  0,\pi\right)  $, while applying the current
removes the degeneracy between them. However, the energy barrier exists for
the switch from one minimum into another. This barrier may disappear if
$\Gamma>1$ and the current is large enough $I>I_{c}/\Gamma$. \ In this regime
the superconducting current would provoke the switching of the magnetization
between one stable configuration $\theta=0$ and another $\theta=\pi$. This
corresponds to the transitions of the junction between $\ +\varphi_{0}$ and
$-\varphi_{0}$ states. The read-out of the state of the $\varphi_{0}$-junction
may be easily performed if it is a part of some SQUID-like circuit (the
$\varphi_{0}$-junction induces a shift of the diffraction pattern by
$\varphi_{0}$).

\begin{figure*}[ptb]
\begin{center}
\includegraphics[width=5.0in]{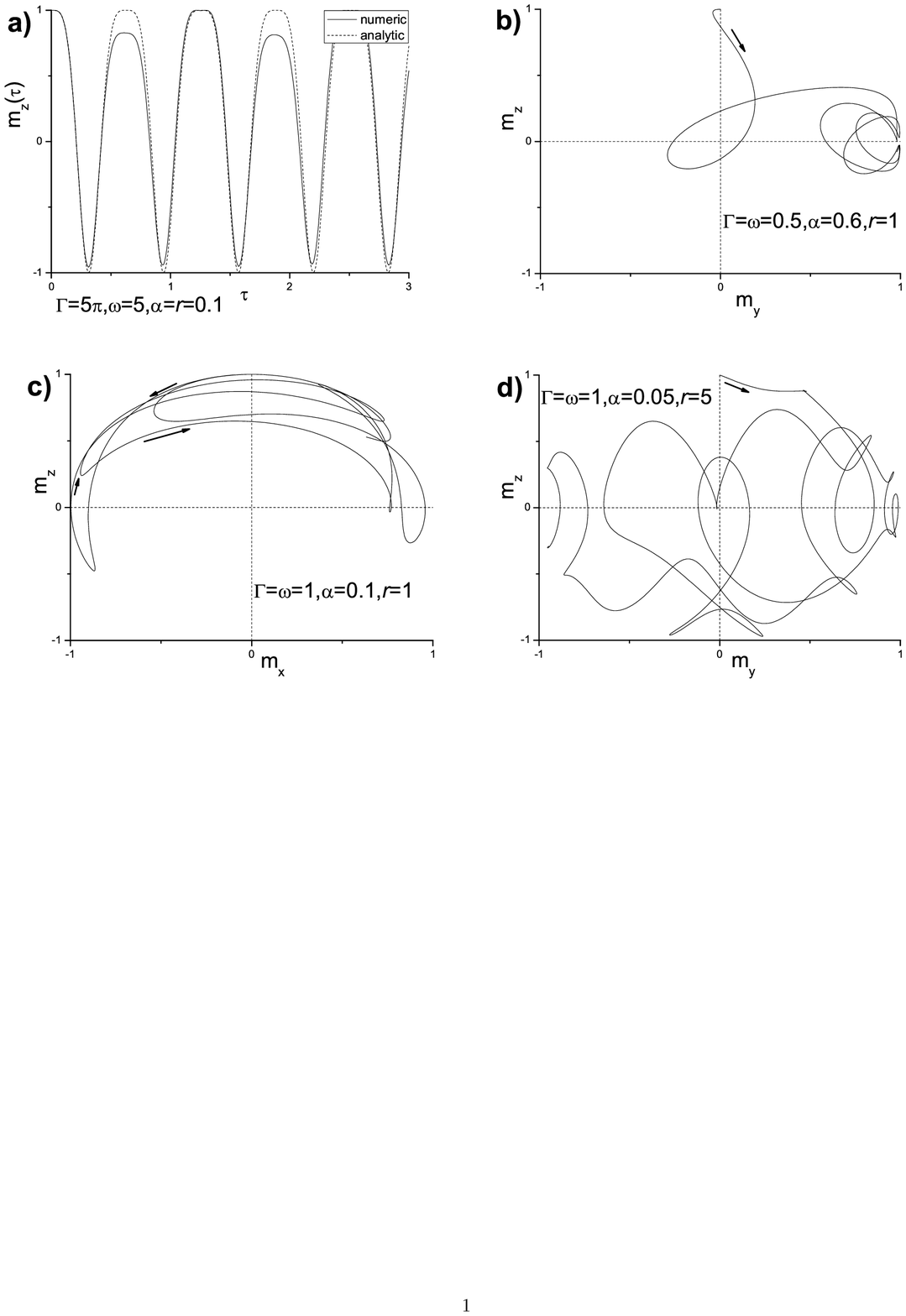}
\end{center}
\caption{Results of numerical analysis of the magnetic moment dynamics of the
$\varphi_{0}$-junction.\ a) Reversal of $m_{z}$ from analytical expression
Eq.$\left(  \ref{EQ_reversal}\right)  $ (dashed curve) and numerical one
(normal curve). The other curves are related to the $\mathbf{M}$ trajectory:
b) in strong damping case c) and d) in the strong coupling regime revealing
incomplete and complete magnetic moment reversal, respectively.}%
\label{FIG_numerical}%
\end{figure*}

In fact, the voltage-biased Josephson junction, and thus the a.c. Josephson
effect provides an ideal tool to study magnetic dynamics in a $\varphi_{0}%
$-junction. In such a case, the superconducting phase varies with time like
$\varphi\left(  t\right)  =\omega_{J}t$ \cite{b.josephson.1968}. If
$\hslash\omega_{J}\ll T_{c}$, one can use the static value for the energy of
the junction $\left(  \ref{EQ_energy}\right)  $ considering $\varphi\left(
t\right)  $ as an external potential. The magnetization dynamics are described
by the Landau-Lifshitz-Gilbert equation (LLG) \cite{b.landau.IX_e}%
\begin{equation}
\frac{d\mathbf{M}}{dt}=\gamma\mathbf{M\times H}_{\text{eff}}+\frac{\alpha
}{M_{0}}\left(  \mathbf{M\times}\frac{d\mathbf{M}}{dt}\right)  ,
\label{EQ_LLG}%
\end{equation}
where $\mathbf{H}_{\text{eff}}=-\delta F/\mathcal{V}\delta\mathbf{M}$ is the
effective magnetic field applied to the compound, $\gamma$ the gyromagnetic
ratio, and $\alpha$ a phenomenological damping constant. The corresponding
free energy $F=E_{s}+E_{M}$ yields
\begin{equation}
\mathbf{H}_{\text{eff}}=\frac{K}{M_{0}}\left[  \Gamma\sin\left(  \omega
_{J}t-r\dfrac{M_{y}}{M_{0}}\right)  \boldsymbol{\hat{y}}+\dfrac{M_{z}}{M_{0}%
}\boldsymbol{\hat{z}}\right]  , \label{EQ_Heff}%
\end{equation}
where $r=\ell v_{\text{so}}/v_{F}$. Introducing $m_{i}=M_{i}/M_{0}$,
$\tau=\omega_{F}t$ ($\omega_{F}=\gamma K/M_{0}^{2}$ is the frequency of the
ferromagnetic resonance) in LLG equation $\left(  \ref{EQ_LLG}\right)  $ leads
to
\begin{equation}
\left\{
\begin{array}
[c]{l}%
\dot{m}_{x}=m_{z}\left(  \tau\right)  m_{y}\left(  \tau\right)  -\Gamma
m_{z}\left(  \tau\right)  \sin\left(  \omega\tau-rm_{y}\right) \\
\dot{m}_{y}=-m_{z}\left(  \tau\right)  m_{x}\left(  \tau\right) \\
\dot{m}_{z}=\Gamma m_{x}\left(  \tau\right)  \sin\left(  \omega\tau
-rm_{y}\right)
\end{array}
\right.  , \label{EQ_LLG_complete}%
\end{equation}
where $\omega=\omega_{J}/\omega_{F}$. The generalization of Eq.$\left(
\ref{EQ_LLG_complete}\right)  $ for $\alpha\neq0$ is straightforward.\ One
considers first the "weak coupling" regime $\Gamma\ll1$ when the Josephson
energy $E_{J}$ is small in comparison with the magnetic energy $E_{M}$. In
this case, the magnetic moment precess around the $z$-axis. If the other
components verify $\left(  m_{x},m_{y}\right)  \ll1$, then the equations
$\left(  \ref{EQ_LLG_complete}\right)  $ may be linearized, and the
corresponding solutions are
\begin{equation}
m_{x}\left(  t\right)  =\dfrac{\Gamma\omega\cos\omega_{J}t}{1-\omega^{2}%
}~\text{and}~m_{y}\left(  t\right)  =-\dfrac{\Gamma\sin\omega_{J}t}%
{1-\omega^{2}}. \label{EQ_magnetic_resonance_without_damping}%
\end{equation}
Near the resonance $\omega_{J}\approx\omega_{F}$, the conditions of
linearization are violated and it is necessary to take the damping into
account.\ The precessing magnetic moment influences the current through the
$\varphi_{0}$-junction like
\begin{equation}
\frac{I}{I_{c}}=\sin\omega_{J}t+\frac{\Gamma r}{2}\dfrac{1}{\omega^{2}-1}%
\sin2\omega_{J}t+..., \label{EQ_current_without_damping}%
\end{equation}
\emph{i.e.}, in addition to the first harmonic oscillations, the current
reveals higher harmonics contributions. The amplitude of the harmonics
increases near the resonance and changes its sign when $\omega_{J}=\omega_{F}%
$. Thus, monitoring the second harmonic oscillations of the current would
reveal the dynamics of the magnetic system.\ 

The damping plays an important role in the dynamics of the considered system.
It results in a d.c. contribution to the Josephson current. Indeed, the
corresponding expression for $m_{y}\left(  t\right)  $ in the presence of
damping becomes
\begin{equation}
m_{y}\left(  t\right)  =\frac{\omega_{+}-\omega_{-}}{r}\sin\omega_{J}%
t+\frac{\alpha_{-}-\alpha_{+}}{r}\cos\omega_{J}t,
\label{EQ_magnetic_resonance_with_damping}%
\end{equation}
where%
\begin{equation}
\omega_{\pm}=\dfrac{\Gamma r}{2}\dfrac{\omega\pm1}{\Omega_{\pm}}\text{ and
}\alpha_{\pm}=\dfrac{\Gamma r}{2}\dfrac{\alpha}{\Omega_{\pm}},
\end{equation}
with $\Omega_{\pm}=\left(  \omega\pm1\right)  ^{2}+\alpha^{2}$. It thus
exhibits a damped resonance as the Josephson frequency is tuned to the
ferromagnetic one $\omega\rightarrow1$. Moreover, the damping leads to the
appearance of out of phase oscillations of $m_{y}\left(  t\right)  $ (term
proportional to $\cos\omega_{J}t$ in Eq.$\left(
\ref{EQ_magnetic_resonance_with_damping}\right)  $). In the result the
current
\begin{multline}
I\left(  t\right)  \approx I_{c}\sin\omega_{J}t+I_{c}\frac{\omega_{+}%
-\omega_{-}}{2}\sin2\omega_{J}t+\\
+I_{c}\frac{\alpha_{-}-\alpha_{+}}{2}\cos2\omega_{J}t+I_{0}\left(
\alpha\right)  \label{EQ_current_with_damping}%
\end{multline}
acquires a d.c. component
\begin{equation}
I_{0}\left(  \alpha\right)  =\frac{\alpha\Gamma r}{4}\left(  \frac{1}%
{\Omega_{-}}-\frac{1}{\Omega_{+}}\right)  .
\end{equation}
This d.c. current in the presence of a constant voltage $V$ applied to the
junction means a dissipative regime which can be easily detected. In some
aspect, the peak of d.c. current near the resonance is reminiscent of the
Shapiro steps effect in Josephson junctions under external r.f. fields. Note
that the presence of the second harmonic in $I\left(  t\right)  $ Eq.$\left(
\ref{EQ_current_with_damping}\right)  $ should also lead to half-integer
Shapiro steps in $\varphi_{0}$-junctions
\cite{sellier_baraduc_lefloch_calemczuk.2004}.

The limit of the "strong coupling" $\Gamma\gg1$ (but $r\ll1$) can also be
treated analytically. In this case, $m_{y}\approx0$ and solutions of
Eq.$\left(  \ref{EQ_LLG_complete}\right)  $ yields
\begin{equation}
\left\{
\begin{array}
[c]{l}%
m_{x}\left(  t\right)  =\sin\left[  \dfrac{\Gamma}{\omega}\left(  1-\cos
\omega_{J}t\right)  \right] \\
\\
m_{z}\left(  t\right)  =\cos\left[  \dfrac{\Gamma}{\omega}\left(  1-\cos
\omega_{J}t\right)  \right]
\end{array}
\right.  , \label{EQ_reversal}%
\end{equation}
which are the equations of the magnetization reversal, a complete reversal
being induced by $\Gamma/\omega>\pi/2$. Strictly speaking, these solutions are
not exact oscillatory functions in the sense that $m_{z}\left(  t\right)  $
turns around the sphere center counterclockwise before reversing its rotation,
and returns to the position $m_{z}\left(  t=0\right)  =1$ clockwise, like a
pendulum in a spherical potential (see Fig.\ref{FIG_numerical}.c).

Finally, we have performed numerical studies of the non-linear LLG Eq.$\left(
\ref{EQ_LLG}\right)  $ for some choices of the parameters when the analytical
approaches fail. To check the consistency of our numerical and analytical
approaches, we present in Fig.\ref{FIG_numerical}.a) the corresponding
$m_{z}\left(  t\right)  $ dependences for low-damping regimes. They clearly
demonstrate the possibility of the magnetization reversal. In
Figs.\ref{FIG_numerical}b-d), some trajectories of the magnetization vectors
are presented for general coupling regimes. These results demonstrate that the
magnetic dynamics of S/F/S $\varphi_{0}$-junction may be pretty complicated
and strongly non-harmonic.

If the $\varphi_{0}$-junction is exposed to a microwave radiation at angular
frequency $\omega_{1}$, the physics that emerge are very rich. First, in
addition to the Shapiro steps at $\omega_{J}=n\omega_{1}$, half-integer-steps
will appear. Secondly, the microwave magnetic field may also generate an
additional magnetic precession with $\omega_{1}$ frequency. Depending on the
parameters of $\varphi_{0}$-junction and the amplitude of the microwave
radiation the main precession mechanism may be related either to the Josephson
current or the microwave radiation. In the last case the magnetic spin-orbit
coupling may substantially contribute to the amplitude of the Shapiro steps.
Therefore, we could expect a dramatic increase of this amplitude at
frequencies near the ferromagnetic resonance. When the influence of the
microwave radiation and Josephson current on the precession is comparable, a
very complicated regime may be observed.

In the present work we considered the case of the easy-axis magnetic
anisotropy. If the ferromagnet presents an easy-plane anisotropy than
qualitatively the main conclusions of this article remain the same because the
coupling between magnetism and superconductivity depends only on the $M_{y}$
component.\ However, the detailed dynamics would be strongly affected by a
weak in-plane anisotropy.\ 

To summarize, we have demonstrated that S/F/S $\varphi_{0}$-junctions provide
the possibility to generate magnetic moment precession via Josephson current.
In the regime of strong coupling between magnetization and current, magnetic
reversal may also occur. These effects have been studied analytically and
numerically. We believe that the discussed properties of the $\varphi_{0}%
$-junctions could open interesting perspectives for its applications in spintronics.\ 

The authors are grateful to Z. Nussinov, J. Cayssol, M. Houzet, D. Gusakova,
M. Roche and D. Braithwaite for useful discussions and comments. This work was
supported by the French ANR Grant N$%
{{}^\circ}%
$ ANR-07-NANO-011: ELEC-EPR.

\end{document}